\documentclass[twocolumn,showpacs,preprintnumbers,amsmath,amssymb]{revtex4}
\usepackage{amssymb}
\usepackage{graphicx,color}
\usepackage{bm}

\begin{document}

\title{Tomography of high-energy nuclear collisions with photon-hadron correlations }

\author{Hanzhong Zhang$^{1,2}$, J. F. Owens$^3$, Enke Wang$^{1,2}$ and Xin-Nian Wang$^{4,1}$}

\affiliation{$^1$Institute of Particle Physics, Huazhong Normal
University,
         Wuhan 430079, China\\
         $^2$Key Laboratory of Quark $\&$ Lepton Physics (Huzhong Normal
University), Ministry of Education, China\\
         $^3$Physics Department, Florida State University, Tallahassee,
          FL 32306-4350, USA\\
$^4$Nuclear Science Division, Lawrence Berkeley Laboratory,
         Berkeley, California 94720, USA}

\date{\today}

\begin{abstract}
Within the next-to-leading order (NLO) perturbative QCD (pQCD)
parton model, suppression of away-side hadron spectra
associated with a high $p_{T}$ photon due to parton energy loss is studied
in high-energy heavy-ion collisions. Dictated
by the shape of the $\gamma$-associated jet spectrum in NLO pQCD,
hadron spectra at large $z_T=p_T^{h}/p_T^{\gamma} \stackrel{>}{\sim}
1$ are more sensitive to parton energy loss and therefore are dominated
by surface emission of $\gamma$-associated jets, whereas small
$z_{T}$ hadrons mainly come from fragmentation of jets with reduced
energy which is controlled by the volume emission. These lead to
different centrality dependence of the $\gamma$-hadron suppression for
different values of $z_{T}$. Therefore, a complete measurement of
the suppression of $\gamma$-triggered hadron spectra, 
including its dependence on the orientation of the $\gamma$-hadron pair 
with respect to the reaction plane,
allows the extraction of the spatial distribution of jet quenching
parameters, achieving a true tomographic study of the quark-gluon
plasma in high-energy heavy-ion collisions.

 \end{abstract}


\maketitle


Jet quenching \cite{wg90} or suppression of large $p_{T}$ 
hadrons, caused by parton energy loss due to strong
interaction between jets and the dense medium, has become a powerful
tool for the study of properties of the quark-gluon plasma
\cite{review} in  high-energy heavy-ion collisions. Experimental 
studies of jet quenching include suppression of single \cite{phenix,star0},
dihadron (back-to-back) \cite{star} and $\gamma$-hadron spectra. 
The strong suppression of large $p_{T}$ single hadron spectra and
its centrality dependence at the Relativistic Heavy-ion Collider (RHIC) \cite{phenix,star},
indicate a picture of surface emission of jets. Most of jets, initially
produced at the center of collisions, undergo large amount of energy
loss and do not contribute to the final observed large $p_{T}$
hadron spectra. High $p_T$ dihadrons, on the other hand, come not
only from jet pairs close and tangential to the surface of the dense
medium but also from punch-through jets originating from the center
of the system \cite{zoww07}, leading to their increased sensitivity to the 
initial gluon density as compared to single hadrons.

We will focus in this paper on the study of $\gamma$-triggered
away-side hadron spectra in heavy-ion collisions within the next-to-leading order (NLO)
perturbative QCD (pQCD) parton model. Since a photon does not interact
in QCD with the dense medium, its energy approximately reflects that
of the initial jet in $\gamma$-jet events prior to jet propagation through 
the medium. One can therefore study the medium
modification of the full jet fragmentation function (FF)
\cite{whs96,renk}. By selecting $\gamma$-hadron pairs with different
values of  $z_{T}=p_{T}^{h}/p_{T}^{\gamma}$, which could be larger than 1
due to radiative correction in NLO pQCD, one can effectively control
hadron emission from different regions of the dense medium and
therefore extract the corresponding jet quenching parameters.

Within pQCD parton model, the NLO corrections ($\alpha_e\alpha_s^2$)
to photon and photon-hadron production cross sections include 1-loop corrections to
$2\rightarrow2$ tree level processes and $2\rightarrow 3$ tree level
contributions which have two-cutoff parameters, $\delta_s$ and
$\delta_c$, to isolate the soft and collinear divergences in the
squared matrix elements \cite{owns1,owens87-90}. This results in a set of
two-body and three-body weights that depend on the cut-offs. The dependence 
cancels when the weights are combined in the calculation
of physical observable, such as inclusive photon and
photon-hadron cross sections. In this paper, we will use a Monte Carlo
implementation \cite{owens87-90} of the NLO pQCD calculation for the
invariant cross section of photon and photon-hadron production.

Photon production in pQCD includes direct photons from the
Compton and annihilation subprocesses and bremsstrahlung photons
from jet fragmentation in high-energy $p+p$ collisions. Since the
bremsstrahlung photons are always accompanied by nearly collinear
hadrons, an isolation-cut can be applied on the photon signal to
suppress bremsstrahlung-like photons both in theory
\cite{owens87-90} and experiments
\cite{phenix-iso-pho}. Even though such isolation-cut cannot
eliminate fragmentation contributions, it can significantly reduce
the fraction of fragmentation photons to about less than 10\% for
$p_{T}^{\gamma}>7$ GeV/$c$. For the study of $\gamma$-hadron
correlation in this paper, we will focus mainly on photons
with isolation cuts. Therefore, we can neglect those photons that
are produced via induced bremsstrahlung \cite{vitevzhang}, jet-photon
conversion \cite{Fries} and thermal production
\cite{Srivastava,TGJM} in  high-energy heavy-ion collisions.

To take into account parton energy loss in $\gamma$-hadron correlation
in high-energy heavy-ion collisions, we will use an effective form
of medium modified FF's as in previous studies \cite{zoww07,Wang:2003mm}
with an average energy loss,
\begin{equation}
\Delta E \approx \langle \frac{dE}{dL}\rangle_{1d}
\int_{\tau_0}^{\infty} d\tau \frac{\tau-\tau_0}{\tau_0\rho_0}
\rho_g(\tau,{\bf b},{\bf r}+{\bf n}\tau),
\end{equation}
 for a quark (gluon energy loss is 9/4 of a quark) produced at a
position ${\bf r}$ and traveling along the direction ${\bf n}$, where,
\begin{equation}
 \langle\frac{dE}{dL}\rangle_{1d}=\epsilon_{0} (E/\mu_0-1.6)^{1.2}
 /(7.5+E/\mu_0)
\label{eq:loss}
\end{equation}
is the parameterized ~\cite{ww01} average quark energy loss per unit length
in a 1-d expanding medium with an initial average gluon density $\rho_0$ at
time $\tau_0$.  The parameter $\epsilon_{0}$ is the initial quark energy
loss per unit distance and is proportional to $\rho_{0}$.
A hard-sphere nuclear overlap geometry is used as in previous studies \cite{zoww07,Wang:2003mm}, 
which differs at most about 10\% from a Wood-Saxon geometry.

We will use the KKP parameterization \cite{KKP} for parton FF's in vacuum.
The parton distributions per nucleon  inside a nucleus are assumed
to be factorizable into parton distributions in a nucleon given
by the CTEQ6M parameterization \cite{CTEQ} and a nuclear modification
factor \cite{hijing}, including the isospin dependence. In the
following, we will use values of the energy loss parameters
($\epsilon_{0}=1.68$ GeV/fm, $\mu_{0}=1.5$ GeV and $\tau_{0}=0.2$ fm/$c$)
as determined from a {\it simultaneous} fit to the suppression of both single and
dihadron spectra \cite{zoww07}, except when specified.

For the study of $\gamma$-hadron spectra we follow Ref.~\cite{whs96} and
define the $\gamma$-triggered FF as,
\begin{eqnarray}
  D_{AA}(z_T)&\equiv&
  \frac{\int d\phi dp_T^{\gamma}dy^{\gamma} dy^{h}
 p_T^{\gamma} \frac{d\sigma^{\gamma h}_{AA}}{dp^{\gamma}_Tdy^{\gamma} dp_T^{h}dy^{h}d\phi}}
  {\int dp^{\gamma}_Tdy^{\gamma} \frac{d\sigma^{\gamma}_{AA}}{dp^{\gamma}_Tdy^{\gamma}}}\,,
  \label{Daa_zt}
\end{eqnarray}
where $z_T=p^{h}_T/p^{\gamma}_T$ and $-\pi/2<\phi<\pi/2$ is
the azimuthal angle between the triggered $\gamma$ and the associated
hadron on the away-side. The above $\gamma$-triggered FF is 
a sum of FF's of the away-side jets (quark and gluon) weighted with
the fractional $\gamma$-jet production cross sections  and convoluted with the transverse momentum
smearing due to NLO processes, which can be exploited to explore
different limit of jet quenching as we will discuss later. The hadron-triggered FF's 
have been similarly defined for the study of dihadron spectra \cite{zoww07,Wang:2003mm} by
replacing photons with triggered hadrons in the above equation.

\begin{figure}
\begin{center}
\includegraphics[width=90mm]{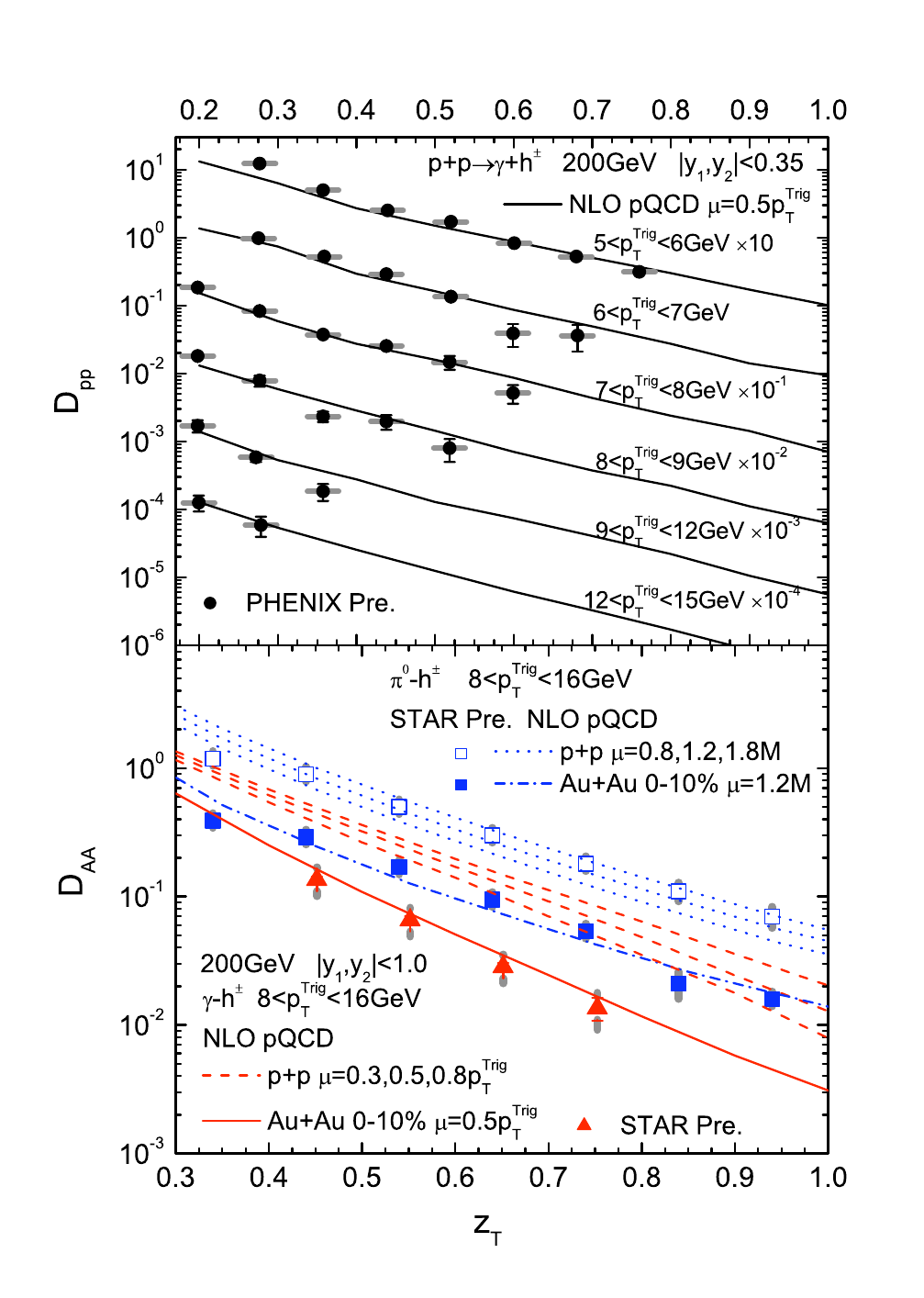}
\end{center}
\vspace{-12pt} \caption{\label{fig:Daa} (color online). $\gamma$-triggered and
hadron-triggered FF 's in $p+p$ and central
$Au+Au$ collisions at the RHIC energy. The preliminary data are from \cite{star-gam-hadr,frantz}.
Systematic errors of experimental data are shown as shaded bars when available.}
\end{figure}

Shown in the upper panel of Fig.~\ref{fig:Daa} are the calculated $\gamma$-triggered FF's in $p+p$
collisions at the RHIC energy as compared to PHENIX preliminary data. They agree very well for 
different values of $p_{T}^{\rm trig}$. In the lower panel of Fig.~\ref{fig:Daa}, we show the $\gamma$-triggered 
FF in central $Au+Au$ (solid) for $8<p_{T}^{\rm trig}<16$ GeV/$c$ as compared to
$p+p$ (dashed) collisions.  With the same energy loss parameters as determined by the 
single and dihadron suppression \cite{zoww07}, the NLO pQCD
results agree with the STAR preliminary data very well. Also shown are the calculated
hadron-triggered FF's as compared to the experimental data, which are
larger than the $\gamma$-triggered FF's. This is mainly because the  fraction of gluon jets 
associated with a hadron trigger at this range of $p_{T}^{\rm trig}$ is larger
than $\gamma$-triggered jets and the hadron yield of gluon jets is larger than that of quarks.
Note that the average initial jet energy associated with a trigger hadron
is larger than that associated with a direct photon for the same value of $p_{T}^{\rm trig}$
due to parton fragmentation.
We also show in the lower panel the uncertainty, mainly from the scale dependence of the FF's, 
of NLO pQCD results for $p+p$ due to the choice of factorization scale $\mu$. 
Here $M$ is the invariant mass of the dihadron. We will use $\mu=0.5p_{T}^{\rm trig}$ 
for $\gamma$-hadron spectra in this paper unless specified.

\begin{figure}
\begin{center}
\includegraphics[width=93mm]{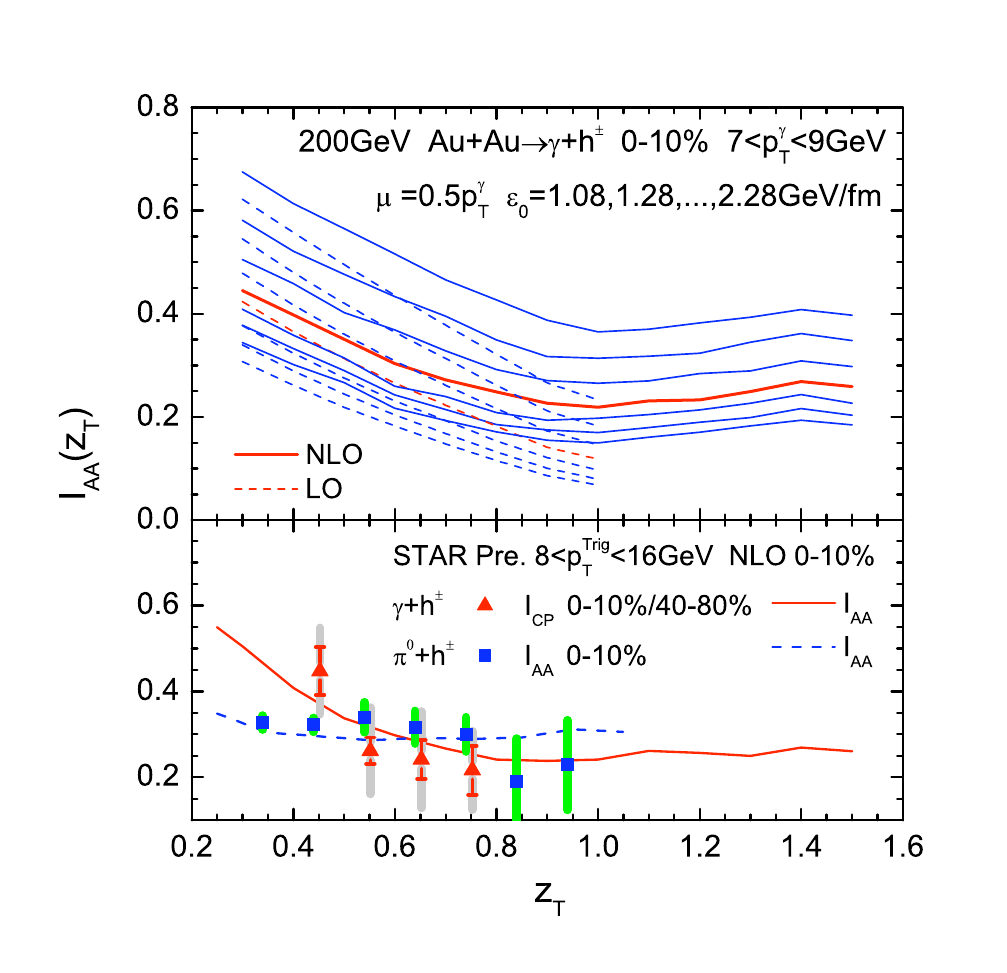}
\end{center}
\vspace{-12pt} \caption{\label{fig:Iaa-zt} (color online). (upper panel) LO and NLO pQCD results
for the suppression factor for $\gamma$-triggered hadrons in
central $Au+Au$ collisions at RHIC energy with different values of energy loss
parameter $\epsilon_{0}$. (lower panel) NLO results for $\gamma$-triggered hadrons
are compared to dihadron suppression together with STAR data \cite{star-gam-hadr}.}
\end{figure}

To quantify the suppression of hadron and $\gamma$-triggered FF's
in central $Au+Au$ relative to $p+p$ collisions due to jet quenching,
as seen in Fig.~\ref{fig:Daa}, one defines the nuclear modification or suppression factor,
\begin{equation}
 I_{AA}={D_{AA}}/{D_{pp}},
\end{equation}
for both hadron and $\gamma$-triggered FF.
Shown in  the upper panel of Fig.~\ref{fig:Iaa-zt} are the calculated nuclear
modification factors both in LO (dashed) and NLO (solid) for
different values of the energy loss parameter $\epsilon_{0}$.  As
compared to the same study of the hadron-triggered FF \cite{zoww07}, 
$I_{AA}(z_{T})$ for $\gamma$-triggered hadron spectra at small $z_{T}\le 0.6$, 
which are dominated by volume emission, is more sensitive to the 
energy loss parameter $\epsilon_{0}$ and therefore provides a better 
phenomenological constraint on the medium properties
when compared to experimental data.  

In the LO pQCD calculation, transverse momentum of the associated
jet is balanced exactly by the direct photon in tree $2 \rightarrow
2$ processes. This limits $z_{T}=p_{T}^{h}/p_{T}^{\gamma}\leq 1$. In
NLO, however,  the initial jet energy can
exceed $p_{T}^{\gamma}$ due to radiative correction or
broadening in the initial state and therefore leads to hadrons with
 $p_{T}^{h}>p_{T}^{\gamma}$ or $z_T>1$. In
this region, the $\gamma$-triggered FF is mainly
determined by the tail of the radiative broadening which falls
sharply as a function of the jet transverse momentum. Therefore,
contributions to the final associated hadron spectra $D_{AA}(z_{T})$
($z_{T}> 1$) from $\gamma$-triggered jets with
even a small amount of energy loss will be suppressed. Only those
jets from surface emission that escape from the medium without energy loss
will contribute, 
whose FF's are the same as in the vacuum. Therefore,
the nuclear modification factor $I_{AA}(z_{T})$ in this region is mainly determined by
the thickness of the corona of the surface emission. On the other
hand, jets that have lost finite amount of energy before fragmenting
into hadrons will contribute to the $\gamma$-triggered FF 
in the region $z_{T}<0.6$ where the nuclear modification factor
$I_{AA}(z_{T})$ is controlled by volume emission of
jets and is therefore more sensitive to the variation of the energy
loss parameter. The value and $z_{T}$ dependence of $I_{AA}(z_{T})$ in $0.6<z_{T}<1.4$
are determined by the competition of the two mechanisms of hadron
emission. One can, therefore,  determine the jet energy loss parameter 
from different spatial regions in heavy-ion collisions by measuring
the effective $\gamma$-triggered FF in the whole
range of $z_{T}$, possibly also for different orientation of the $\gamma$-hadron pair with 
respect to the reaction plane, achieving a true tomographic
study of the dense medium. For precision studies one should also consider
the effect of intrinsic transverse momentum broadening via systematic analysis
of $p+p$ and $p(d)+A$ collisions \cite{whs96}.

The suppression factors $I_{AA}$ for both hadron (dashed) and $\gamma$-triggered (solid) FF's
in central $Au+Au$ collisions at the RHIC energy are compared
with the STAR data in the lower panel of Fig.~\ref{fig:Iaa-zt}.
The stronger dependence of $I_{AA}$ for $\gamma$-hadrons on $z_{T}$
is due to the dominance of volume emission at small $z_{T}$. 
The similarity in value between $I_{AA}$ for dihadron and $\gamma$-hadron
spectra at $z_{T} \approx 0.4-1.0$, despite of their different trigger biases,
is partially due to the competition between the larger gluon jet fraction (bigger energy loss) 
and larger initial jet energy (more penetration) associated with a hadron trigger. 

\begin{figure}
\begin{center}
\includegraphics[width=85mm]{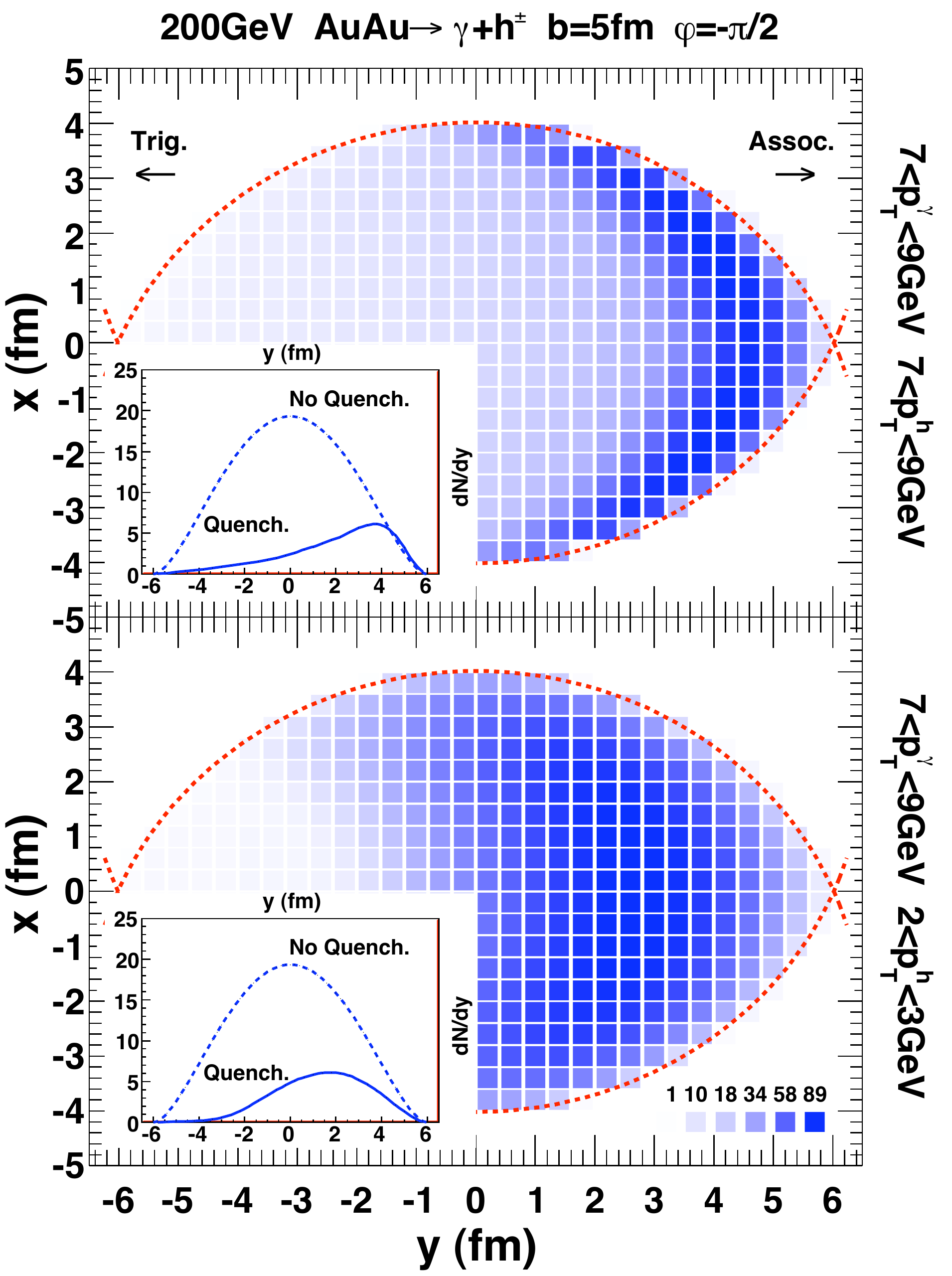}
\end{center}
 \caption{\label{fig:contour} (color online). Transverse spatial distributions
of the initial $\gamma$-jet production vertexes that contribute to
the final observed $\gamma$-hadron pairs along a given direction
(arrows) with $z_T\approx 0.9$ (upper panel) and $z_T\approx 0.3$
(lower panel).}
\end{figure}

To further illustrate the above picture of volume and surface
emission of $\gamma$-triggered jets and their contributions to the
effective $\gamma$-triggered FF
at different values of $z_{T}$, we plot in Fig.~\ref{fig:contour}
the spatial transverse distribution of the initial $\gamma$-jet
production vertexes that contribute to the final $\gamma$-hadron pairs
with given values of $z_{T}$. The color strength represents the
$\gamma$-hadron yield from the fragmentation of the
$\gamma$-triggered jets after parton energy loss.
The $\gamma$-hadron yields with arbitrary scale are given by
Eq.~(\ref{Daa_zt}). The inserted panels are projections of the
contour plots onto $y$-axes with (solid) and without energy
loss (dashed). For $z_{T} \approx 1$ (upper-panel),
contributions to the final observed $\gamma$-hadron pairs indeed 
come mostly from the surface. Contributions
from $\gamma$-triggered jets from the center or the same side of
the trigger photon are mostly suppressed. However, for small
$z_{T}\approx 0.3$ (lower-panel), contributions to the effective
$\gamma$-triggered FF are seen to come from mostly
the whole volume, except near the surface on the side of the
trigger photon.

\begin{figure}
\begin{center}
\includegraphics[width=85mm]{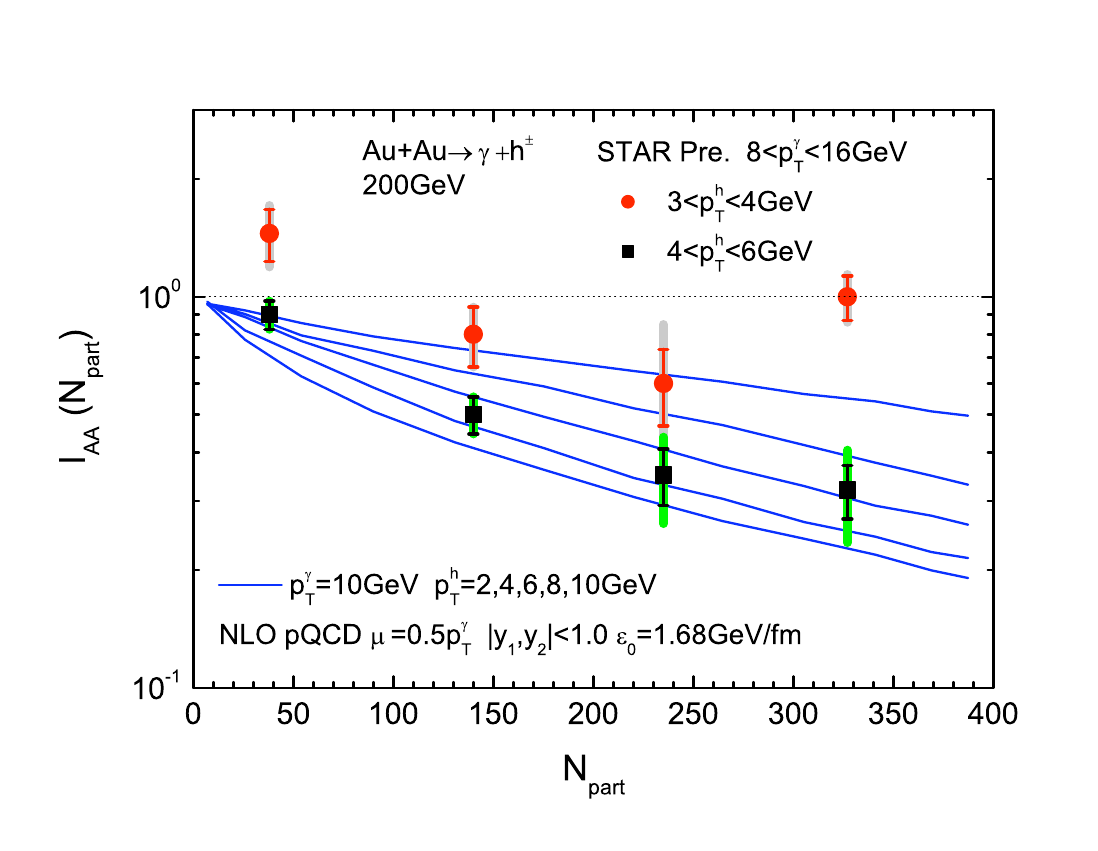}
\end{center}
\vspace{-12pt} \caption{\label{fig:Iaa-Npart-GamHadr} (color online). $N_{\rm part}$ dependence of
$I_{AA}$ for $\gamma$-triggered FF in $Au+Au$ collisions at the RHIC energy. 
The data are from \cite{star-gam-hadr}.  $p_{T}^{h}$ in NLO pQCD calculation is chosen as
$p_T^{h}=2,4,6,8,10$ GeV/$c$, respectively (from top to bottom).}
\end{figure}

These pictures of volume and surface emission of
$\gamma$-hadron pairs in heavy-ion collisions
will lead to different centrality dependence of the suppression
factor $I_{AA}(z_{T})$ for different values of $z_{T}$.
Shown in Fig.~\ref{fig:Iaa-Npart-GamHadr} are $I_{AA}$ 
for $\gamma$-triggered hadron spectra as functions of the
participant number $N_{\rm part}$ in $Au+Au$ collisions at the RHIC energy for
different values of $z_{T}$ as compared to the STAR preliminary
data. For small values of $z_{T}<0.6$, the $\gamma$-triggered hadron
yield is dominated by volume emission and therefore the centrality
dependence of  $I_{AA}$ is weaker than that
in the region $z_{T}\geq 1$ where surface emission is the dominant
production mechanism.

In summary, high $p_T$ $\gamma$-hadron correlations are studied within
the NLO pQCD parton model with modified parton FF 's 
due to jet quenching in high energy heavy-ion collisions. We
demonstrated that volume (surface) emission dominates the
$\gamma$-triggered hadrons spectra at small $z_{T}<0.6$ (large
$z_{T}\geq 1$) due to the underlying jet spectra in the NLO pQCD.
This will enable one to extract jet quenching parameters from
different regions of the dense medium via measurement of the nuclear
modification factor of the $\gamma$-triggered FF
in the whole kinetic region,  including $z_{T}\geq 1$, achieving a
true tomographic study of the dense medium.

We thank P. Jacobs, S. Mioduszewski and M. Nguyen for help comments.
This work was supported by DOE under contracts DE-AC02-05CH11231 and
DEFG02- 97IR40122, by NSFC of China under Projects No. 10825523 and
No. 10875052 and No. 10635020, by MOE of China under Projects No.
IRT0624 and by MOE and SAFEA of China under Project No.
PITDU-B08033.


\begin{thebibliography}{99}

\bibitem{wg90}
X.~N.~Wang and M.~Gyulassy,
  Phys.\ Rev.\ Lett.\  {\bf 68}, 1480 (1992).

\bibitem{review}
  M.~Gyulassy, I.~Vitev, X.~N.~Wang and B.~W.~Zhang
   nucl-th/0302077;
  A.~Kovner and U.~A.~Wiedemann,
  hep-ph/0304151,
   in {\it Quark Gluon Plasma} 3, eds. R. C. Hwa and X.~N. Wang (World Scientific, Singapore, 2003).

\bibitem{phenix} K.~Adcox {\it et al.},
  Phys.\ Rev.\ Lett.\  {\bf 88}, 022301 (2002).

\bibitem{star0}
  C.~Adler {\it et al.}  ,
  Phys.\ Rev.\ Lett.\  {\bf 89}, 202301 (2002).

\bibitem{star} C.~Adler {\it et al.},
  Phys.\ Rev.\ Lett.\  {\bf 90}, 082302 (2003).

\bibitem{zoww07}
  H.~Z.~Zhang, J.~F.~Owens, E.~Wang and X.~N.~Wang,
  Phys.\ Rev.\ Lett.\  {\bf 98}, 212301 (2007).

\bibitem{whs96}
  X.~N.~Wang, Z.~Huang and I.~Sarcevic,
  Phys.\ Rev.\ Lett.\  {\bf 77}, 231 (1996);
  X.~N.~Wang and Z.~Huang,
  Phys.\ Rev.\  C {\bf 55}, 3047 (1997).

\bibitem{renk}
  T.~Renk,
  Phys.\ Rev.\  C {\bf 74}, 034906 (2006).


\bibitem{owns1}
B.~W.~Harris and J.~F.~Owens,
  Phys.\ Rev.\  D {\bf 65}, 094032 (2002).


\bibitem{owens87-90}
H.~Baer, J.~Ohnemus and J.~F.~Owens,
  Phys.\ Rev.\  D {\bf 42}, 61 (1990).


\bibitem{phenix-iso-pho}
S.~S.~Adler {\it et al.},
  Phys.\ Rev.\ Lett.\  {\bf 98}, 012002 (2007).

\bibitem{vitevzhang}
I.~Vitev and B.~W.~Zhang,
  Phys.\ Lett.\  B {\bf 669}, 337 (2008).



\bibitem{Fries}
R.~J.~Fries, B.~Muller and D.~K.~Srivastava,
Phys.\ Rev.\ Lett.\  {\bf 90}, 132301 (2003).


\bibitem{Srivastava}
  D.~K.~Srivastava,
  J.\ Phys.\ G {\bf 35}, 104026 (2008).

\bibitem{TGJM}
S.~Turbide, C.~Gale, S.~Jeon and G.~Moore ,
Phys.\ Rev.\ C.\ {\bf 72}, 014906 (2005).


\bibitem{Wang:2003mm}
  X.~N.~Wang,
  Phys.\ Lett.\ B {\bf 595}, 165 (2004);
  {\bf 579}, 299 (2004).


\bibitem{ww01}
  E.~Wang and X.~N.~Wang,
  Phys.\ Rev.\ Lett.\  {\bf 87}, 142301 (2001).

\bibitem{KKP}B.~A.~Kniehl, G.~Kramer and B.~Potter,
  Nucl.\ Phys.\ B {\bf 582}, 514 (2000).

\bibitem{CTEQ} D.~Stump, J.~Huston, J.~Pumplin, W.~K.~ Tung, H.~L.~Lai, S.~Kuhlmann, J.~Owens,
 JHEP 0310:046(2003).

\bibitem{hijing}
  S.~Y.~Li and X.~N.~Wang,
  Phys.\ Lett.\  B {\bf 527}, 85 (2002).


\bibitem{star-gam-hadr}
J.~Adams {\it et al.} ,
  Phys.\ Rev.\ Lett.\  {\bf 97}, 162301 (2006);
A.~M.~Hamed,
  J.\ Phys.\ G {\bf 35}, 104120 (2008);
  arXiv:0809.1462 [nucl-ex].

\bibitem{frantz}
  J.~Frantz,
  arXiv:0901.1393 [nucl-ex].


\end{thebibliography}
\end{document}